\documentclass[12pt]{iopart}

\usepackage[pdftex]{graphicx}
\graphicspath{{../pdf/}{../jpeg/}}
\DeclareGraphicsExtensions{.pdf,.jpeg,.png}
\usepackage{array}
\expandafter\let\csname equation*\endcsname\relax

\expandafter\let\csname endequation*\endcsname\relax
\usepackage{url}
\usepackage{cite}
\usepackage{chemformula}

\begin{document}

\title[A Broadband and Compact mm-Wave Imaging System based on SAR]{A Broadband and Compact Millimeter-Wave Imaging System based on Synthetic Aperture Radar}

\author{Fahimeh Sepehripour\textsuperscript{1,2}, Ahmad Shafiei Alavijeh\textsuperscript{2},  Mohammad Fakharzadeh\textsuperscript{2,*}, Amin Khavasi\textsuperscript{2}}

\address{1. Eindhoven University of Technology, Eindhoven, Netherlands\\ 2. Sharif University of Technology, Tehran, Iran}
\ead{f.sepehripour@tue.nl, ahmadshafiei\_a@yahoo.com, fakharzadeh@sharif.edu, khavasi@sharif.edu}
\vspace{10pt}
\begin{indented}
\item[]April 2022
\end{indented}

\begin{abstract}
In this paper, the design, realization, and demonstration of a broadband millimeter-wave imaging system based on the  synthetic aperture radar technique (SAR) are discussed. The proposed system, operating within the frequency range of 25.3-30.8\,GHz, consists of a tapered slot antenna as the transmitter and two half-circle antennas as the receivers. The size of the antenna is $19.5\times 8\,mm$ with a maximum gain of $8.5\,dB$. The transmitter and the receiver antennas are printed on the same board. This feature leads to a highly compact and flexible configuration, enabling the applicability of the proposed imaging system in the handheld devices. Furthermore, it significantly reduces the fabrication cost of the system. The proposed broadband imaging system, being capable of performing 3D real-time imaging with high resolutions, can be easily calibrated for each frequency within the desired range. By performing 3D imaging from metallic objects with different shapes, we experimentally demonstrate the high performance of the proposed system, which offers great potentialities for a broad range of applications such as security, medical diagnostic, concealed object detection, to name a few.
\end{abstract}

%
%
%
%
%

\section{Introduction}

Millimeter-wave spectrum, spanning the frequency range of 30-300\,GHz, has many unique properties, making it suitable for a large variety of applications. In particular, the electromagnetic waves within this frequency range can penetrate through the thin layers of dielectric materials \cite{moll2012towards,sheen2001three,lettington2002review,jia2016,patel2016computational,appleby2017millimeter}.  This property allows one to construct the high-resolution image of many materials such as wood, plastic and other dielectric composites that are opaque in the visible range.

Among the different techniques proposed for millimeter-wave  imaging, the so-called synthetic aperture radar (SAR) method \cite{elboushi2012b14,camlica2017,zamani2020random} is one of the most powerful ones, allowing one to generate the image of the object in three dimensions (3D). In this technique, a wideband single element antenna is scanned on a two-dimensional plane located in front of the object. The signal received by the receiver antenna is then processed using the SAR algorithms. The major advantage of SAR is that it allows the construction of 3D images in real-time. Apart from this, SAR enables producing images with a resolution as high as a quarter wavelength \cite{qin2014generalized,ghasr2016wideband}. In addition, in the SAR technique, one can achieve the image of the object by reconstructing only the real part of the hologram. 

To be able to use SAR-based imaging systems in handheld structures, it is important to reduce their size and wight as much as possible. Mono-static millimeter wave imaging systems represent an appropriate platform to reduce the bulkiness of the associated structure, compared to their bi-static and multi-static counterparts. In a mono-static imaging system, one single antenna is used both as the transmitter and the receiver, enabling miniaturization. Despite this advantage, mono-static imaging systems suffer from the difficulties in the calibration process. More specifically, to properly implement a mono-static system, it is required to calibrate the system to a phase reference. For single frequency systems, the calibration can be readily accomplished, since the associated phase error is the same for all points of the synthetic aperture, whereas for a broadband system, the corresponding phase error is not the same for all frequencies. As a result, the calibration must be done for all frequencies individually. This renders the calibration process extremely challenging. Nevertheless, in \cite{ghasr2014novel}, a calibration method is proposed, which can be conveniently performed over a wideband frequency range. In this method, in a single board, the transmitted signal is sampled by a receiver antenna placed at the radiating end of the transmitter antenna. The sampled signal is then used as the reference for all frequencies, mitigating the calibration process. 

In this paper, we demonstrate a wideband millimeter wave imaging system designed based on the SAR technique. The calibration of the proposed system is accomplished by sampling the transmitted signal in the receiver antenna as a reference. Since the receiver and transmitter antennas are printed on the same board, the proposed imaging system is highly compact, enabling portability. The antenna element of the structure is optimally designed such that the transmitter antenna operates in a wide range of frequency with high levels of matching. The proposed millimeter wave imaging system can have a large variety of applications such as nondestructive testing of structures \cite{kempin2013,agarwal2015adaptive,shafi2017,alidoustaghdam2020}, security \cite{appleby2007,sheen2011active,wu2016,zhang2018,briqech202057}, and medical image processing \cite{henriksson2010quantitative,klemm2010,aldhaeebi2019}. 
\section{Millimeter-Wave Imaging Modeling}
\subsection {Hardware Description}
\begin{figure}
  \begin{center}
  \includegraphics[width=3.5in]{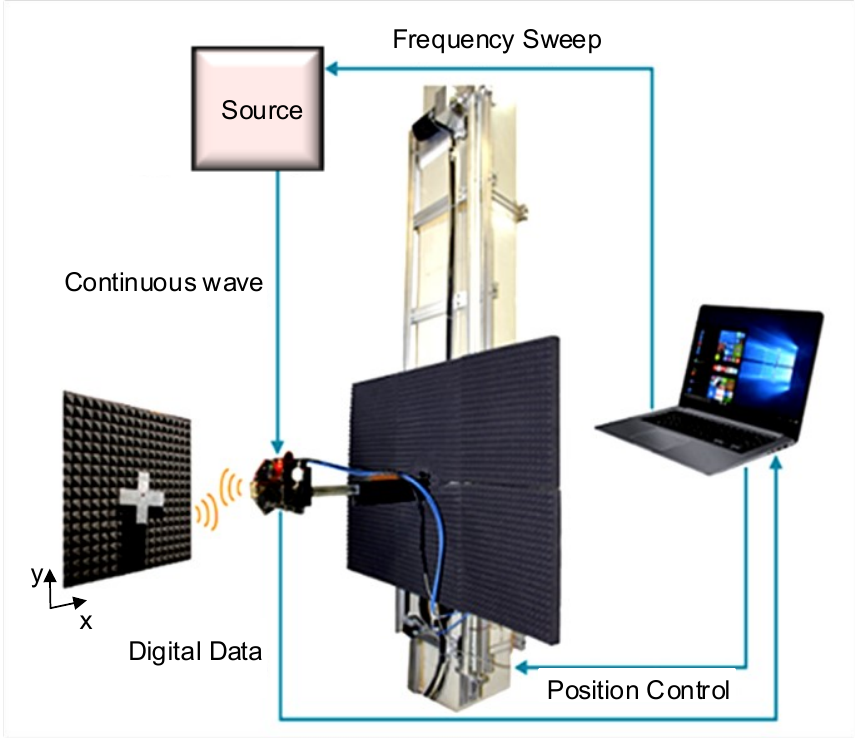}\\
  \caption{Conceptual sketch of the proposed millimeter wave imaging system. The system consists of a probe antenna transmitting and receiving signals to and from an object, a mechanical scanning system, a source, and a data acquisitions unit.}\label{system_model}
  \end{center}
\end{figure}
In this section, we describe the principle of the proposed imaging system. The goal is to realize a millimeter-wave imaging system that reconstructs the 2D or 3D image of a target object. The proposed imaging system is based on sending a millimeter-wave signal to the object, and analyzing the corresponding reflected signal. Fig.~\ref{system_model} illustrates the proposed imaging system. In this system, an RF signal is generated at the frequency range of interest (25.3-30.8\,GHz). The RF signal is then amplified by an amplifier (AMMP6232) with a constant gain of 20 dB in the desired frequency range. The corresponding amplified signal is transformed into a tapered slot antenna, which is the transmitter antenna. The transmitter antenna sends millimeter wave signals to the target object. It is raster-scanned in x-y direction, where the scanning process is controlled by the computer. The corresponding reflected field from the object is received by two receiver diodes. The based-band signals, which are the outputs of the diodes, are then converted into two digitial signals using a control and data acquisition unit. The associated digital signal is then used to reconstruct the image using the generalized synthetic aperture focusing technique (GSAFT) \cite{rezaei2021,zamani20181,qin2014generalized,nilsson2007radar}. 
\subsection{Reconstruction Algorithm}
According to the plane-wave expansion method, the received field by the receiver antennas ($H(x^\prime,y^\prime)$, which is called the hologram) is expressed as \cite{soumekh1991,sheen2001three} 
\begin{equation}\label{eq1}
    H(x',y' )= \iint_S f(x,y) e^{-2jk_0r}dxdy,
\end{equation}
where $(x',y',0)$ and $(x ,y ,z_0)$ correspond to the locations of the antenna and the target, respectively, and $k_0$ is the free space wave-number. Moreover, $f(x,y)$ corresponds to the image of the object. In addition, $r$ is
\begin{equation}\label{eq2}
    r=  \sqrt{(x-x')^2+(y-y')^2+z_0^2}.
\end{equation}
The aim of this imaging system is to reconstruct the image of the object $f(x,y)$. For this purpose,
we decompose the 
exponential term in (\ref{eq1}), which represents a spherical wave, into a superposition of
plane-waves as follows
\begin{equation}\label{eq21}
\begin{split}
     e^{-2jk_0r} =& e^{-2jk_0\sqrt{(x-x')^2+(y-y')^2+z_0^2}}  \\= &\iint e^{jk_{x'}(x'-x)+jk_{y'}(y'-y)+jk_{z}z_0} dk_{x'} dk_{y'},
\end{split}
\end{equation}
where $k_z =\sqrt{4k_0^2 - k_x^2 - k_y^2 }$. By replacing (\ref{eq21}) in (\ref{eq1}), one obtains
\begin{equation}\label{eq22}
\begin{split}
     H(x',y' )= \iint [ \iint & f(x,y)e^{-jk_{x'}x-jk_{y'}y} dxdy]\\
   & e^{jk_{x'}x'+jk_{y'}y'+jk_{z}z_0} dk_{x'} dk_{y'},
\end{split}
\end{equation}
in which the inner double integral represents the 2D Fourier transform of $f(x,y)$, i.e. $\mathcal{F}_{2D}\{f(x,y)\} = F(k_{x'},k_{y'})$. Then 
\begin{equation}\label{eq23}
\begin{split}
     H(x',y' )= \iint F(k_{x'},k_{y'})e^{jk_{x'}x'+jk_{y'}y'+jk_{z}z_0} dk_{x'} dk_{y'},
\end{split}
\end{equation}
by multiplying both sides of (\ref{eq23}) with $e^{-jk_{z}z_0}$, one obtains
\begin{equation}\label{eq24}
\begin{split}
     e^{-jk_{z}z_0}H(x',y' )=& \iint F(k_{x'},k_{y'})e^{jk_{x'}x'+jk_{y'}y'} dk_{x'} dk_{y'} \\
     =&\mathcal{F}_{2D}^{-1}\{\mathcal{F}_{2D}\{f(x,y)\}\}
\end{split}
\end{equation}
$f(x,y)$ can then be expressed according to the following equation
\begin{equation}\label{eq3}
    f(x,y)=\mathcal{F}_{2D}^{-1} \{\mathcal{F}_{2D} \{H(x',y') e^{-jk_z z_0 }   \} \}
\end{equation}
$\mathcal{F}_{2D}$, and $\mathcal{F}_{2D}^{-1}$ denote the 2D Fourier transform and inverse Fourier transform, respectively.

In the proposed system, the outputs of the diodes correspond to the real part of $H(x',y' )$. Here, we prove the possibility of performing imaging using the real part of the hologram. We can rewrite the real part of the hologram as 
\begin{equation}\label{eq4}
    \Re\{H(x',y')\} = \frac{1}{2}(H(x',y')+H^*(x',y'))
\end{equation}
in which $H(x',y')$ is the hologram. Based on the first term on the right side of (\ref{eq4}), one can obtain the image of the object at $z=z_0$ using the following formula \cite{sheen2013sparse}
\begin{equation}\label{eq5}
    f(x,y)=\mathcal{F}_{2D}^{-1} \{\mathcal{F}_{2D} \{H(x',y') e^{-jk_z z_0 }   \} \}
\end{equation}
Likewise, the second term corresponds to the image of the object $z=-z_0$  according to the following equation
\begin{equation}\label{eq6}
    f^*(x,y)=\mathcal{F}_{2D}^{-1} \{\mathcal{F}_{2D} \{H^*(x',y') e^{-jk_z (-z_0) }   \} \}
\end{equation}
This proves that, by using the real part of the hologram, we can reconstruct the image at $z=z_0$. 
\section{Hardware Structure}
\subsection{Transmitter Antenna}
\begin{figure}
  \begin{center}
  \includegraphics[width=3.5in]{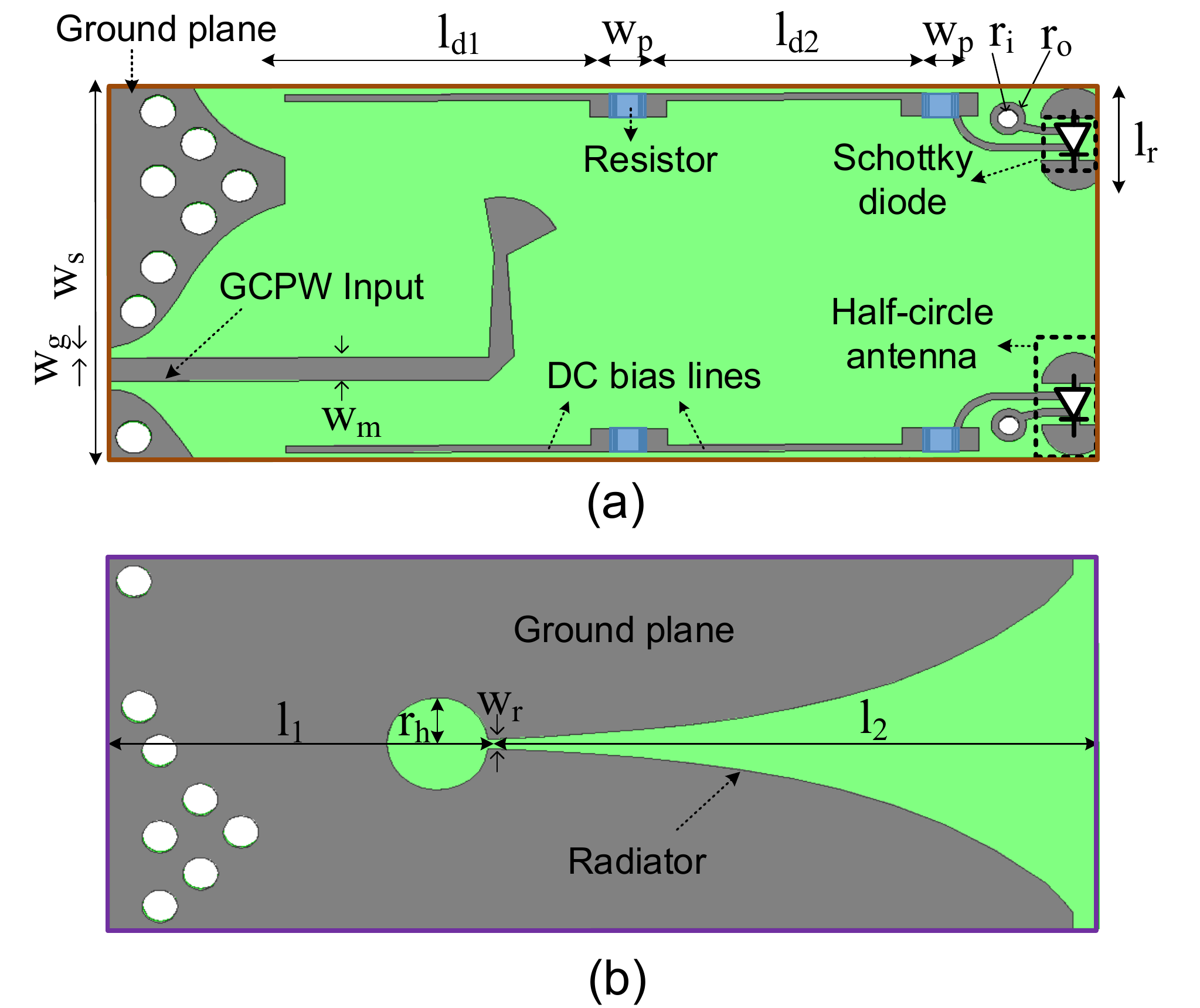}
  \caption{Structure of the designed antenna. (a) Schematic of the top layer of the designed. (b) Bottom layer of the designed antenna. The associated dimensions are summarized in Table 1.}\label{antenna_model}
  \end{center}
\end{figure}

\begin{table}[!t]
\renewcommand{\arraystretch}{1.5}

\caption{Dimensions of the designed Antenna}
\label{table_example}
\centering
\begin{tabular}{|c  ||  c|}
\hline
Parameter & Dimension(mm)\\
\hline
$w_s$ & 8\\
$w_g$ & 0.2\\
$w_m$ & 0.5\\
$w_p$ & 0.5\\
$w_r$ & 0.2\\
$r_i$ & 0.16\\
$r_o$ & 0.31\\
$r_h$ & 1\\
$l_r$ & 2\\
$l_{d1}$ & 10.02\\
$l_{d2}$ & 5.06\\
$l_{1}$ & 7.47\\
$l_{2}$ & 12.03\\
\hline
\end{tabular}
\end{table}

\begin{figure}
  \begin{center}
  \includegraphics[width=3.5in]{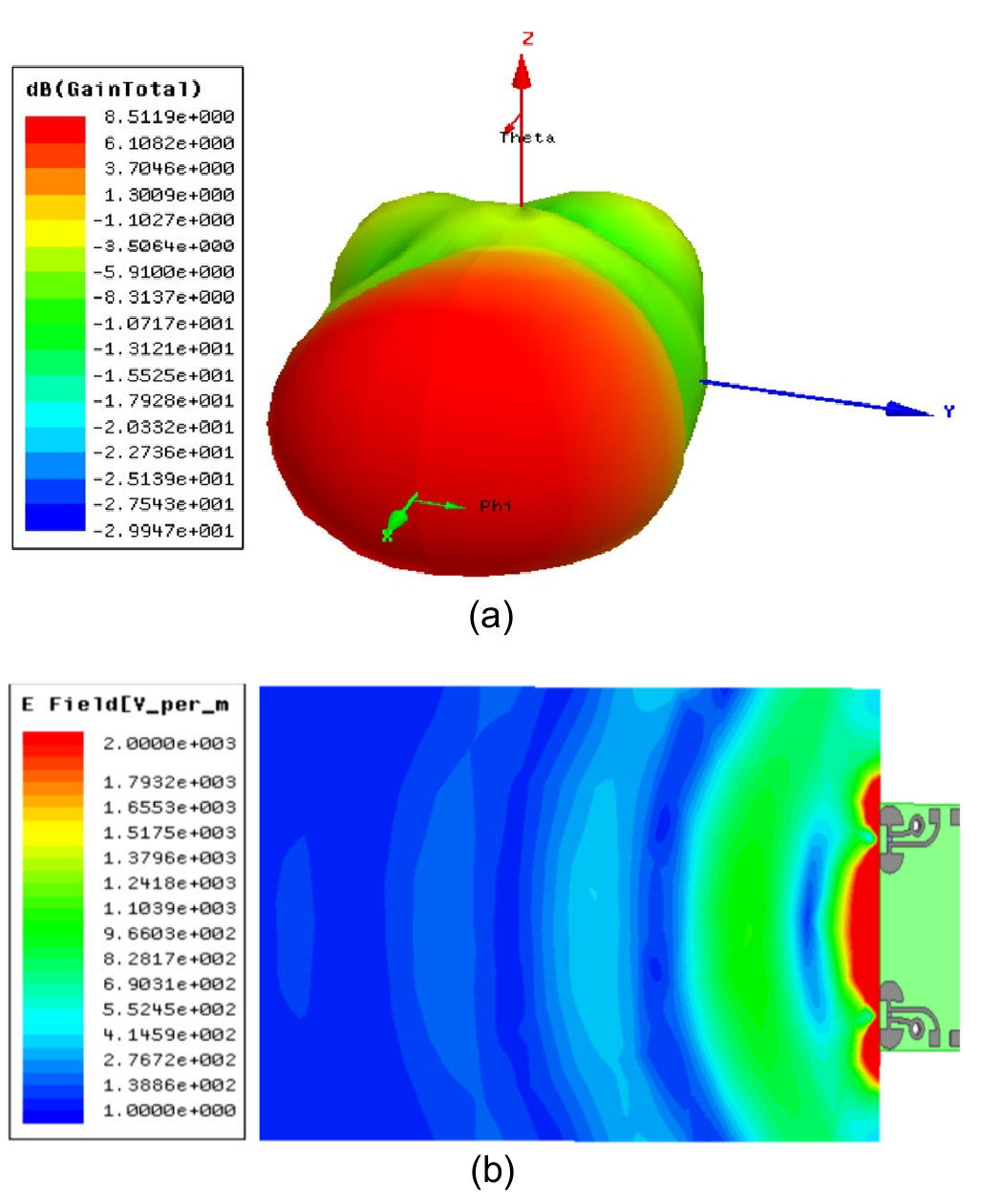}
  \caption{(a) Radiation pattern of the transmitter antenna.  (b) Profile of the electric field. The results are obtained at the frequency of 26\,GHz.}\label{radation_pattern}
  \end{center}
\end{figure}

A key part of a millimeter wave imaging system is the antenna element. The antenna parameters such as the size of the antenna must be designed in a way that the image can be reconstructed using the GSAFT algorithm, which was described before. Fig.~\ref{antenna_model}(a) and (b) show the layouts of the top and bottom layers of the designed antenna element, respectively. The transmitter antenna is a tapered slot antenna. The  tapered slot antenna  brings several advantages \cite{ebnabbasi2012taper,abou2010modulated,sugawara1997mm}. Firstly, it enables a wideband type of operation. In addition, it leads to a relatively high gain, low side-lobe levels, and a symmetrical radiation pattern. The corresponding dimensions of the antenna are summarized in Table I. The associated frequency range is 25.3-30.8\,GHz, which is a desirable region for millimeter wave imaging applications.The operation frequency can be tuned in an arbitrary range by scaling the system dimensions. The three-dimensional radiation pattern of the antenna is shown in Fig.~\ref{radation_pattern}(a), obtained by performing direct finite-element simulations in HFSS at the operation frequency of 26\,GHz. As observed, the antenna has a symmetric radiation pattern, in which the maximum gain ($G_{max} = 8.5\,dB$) happens in the bore-sight direction. Inset of Fig.~\ref{radation_pattern}(b) illustrates how the electric field radiates from the antenna, based on which one can deduce that the antenna has relatively wide beamwidth. Together, these features enable high-resolution imaging based on GSAFT technique.      

\subsection{Receiver}
\begin{figure}
  \begin{center}
  \includegraphics[width=3.5in]{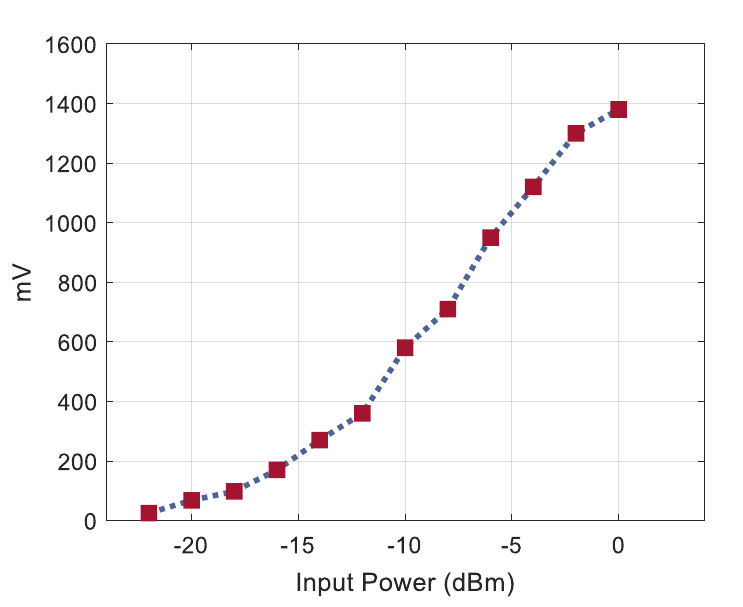}
  \caption{The measured output voltage of the diodes as a function of the input power of the transmitter antenna.  The measurement is performed at f=26\,GHz.}\label{voltage}
  \end{center}
\end{figure}

Two receiver antennas are placed symmetrically at the radiating end of the transmitter, maintaining the radiation pattern of the antenna element symmetric and enabling a denser spatial sampling. Each receiver is composed of a Schottky detector diode (Skyworks SMS7621-060) and a half-circle antenna surrounding the diode. The role of the half-circle antenna is to enhance the coupling to the diode. The reason for choosing half-circle type antenna is its larger electrical length compared to other elements with the same area.  The output of the Schottky diode is a real-valued DC voltage, based on which the phase and the amplitude of the reflected electric field are extracted. This voltage is generated by down-mixing the field scattered from the transmitter and the associated reflected field. To characterize the performance of the diodes, we measured the output voltage of one diode as a function of the input power of the transmitter antenna. The associated measurement results are shown in Fig.~\ref{voltage}.
\subsection{Antenna Element Fabrication}
\begin{figure}
  \begin{center}
  \includegraphics[width=3.5in]{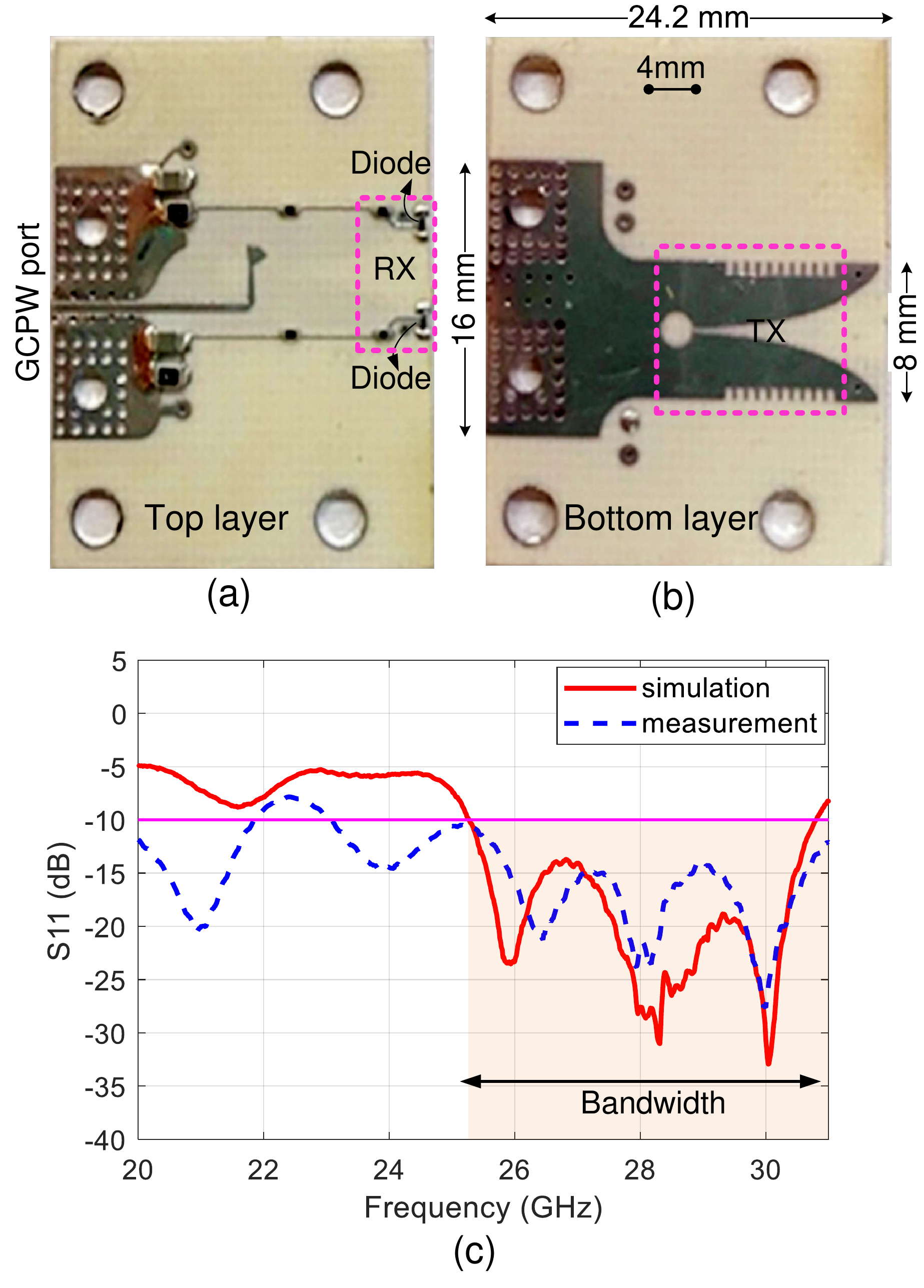}
  \caption{Performance analysis of the fabricated antenna. (a), and (b) Photographs of the top (panel a) and bottom (panel b) of the fabricated antenna. (c) The corresponding measured (blue line) and simulated (red line) $S_{11}$.}\label{fabricated_antenna}
  \end{center}
\end{figure}

Fig.~\ref{fabricated_antenna}(a) and (b) show the photograph of the top and bottom layers of the fabricated antenna. The antenna is printed on a RO4003 Rogers substrate having a thickness of 8 mils. The PCB is fabricated with a minimum gap size of 150 $\mu m$. To excite the antenna, an end-launch connector is used. The associated S parameters are then extracted by performing scattering measurements in an anechoic electromagnetic chamber. Fig.~\ref{fabricated_antenna}(c) represents the measured (dashed blue line) scattering parameter $(S_{11})$ and compares it with simulation (solid red line). The simulation result is obtained by performing the full-wave simulation in Ansys HFSS. As observed, within the frequency range of 25.3-30.8\,GHz, $S_{11}$ of the antenna is below 10 dB (purple line). 

\subsection{Data Acquisition System }
\begin{figure}
  \begin{center}
  \includegraphics[width=3.5in]{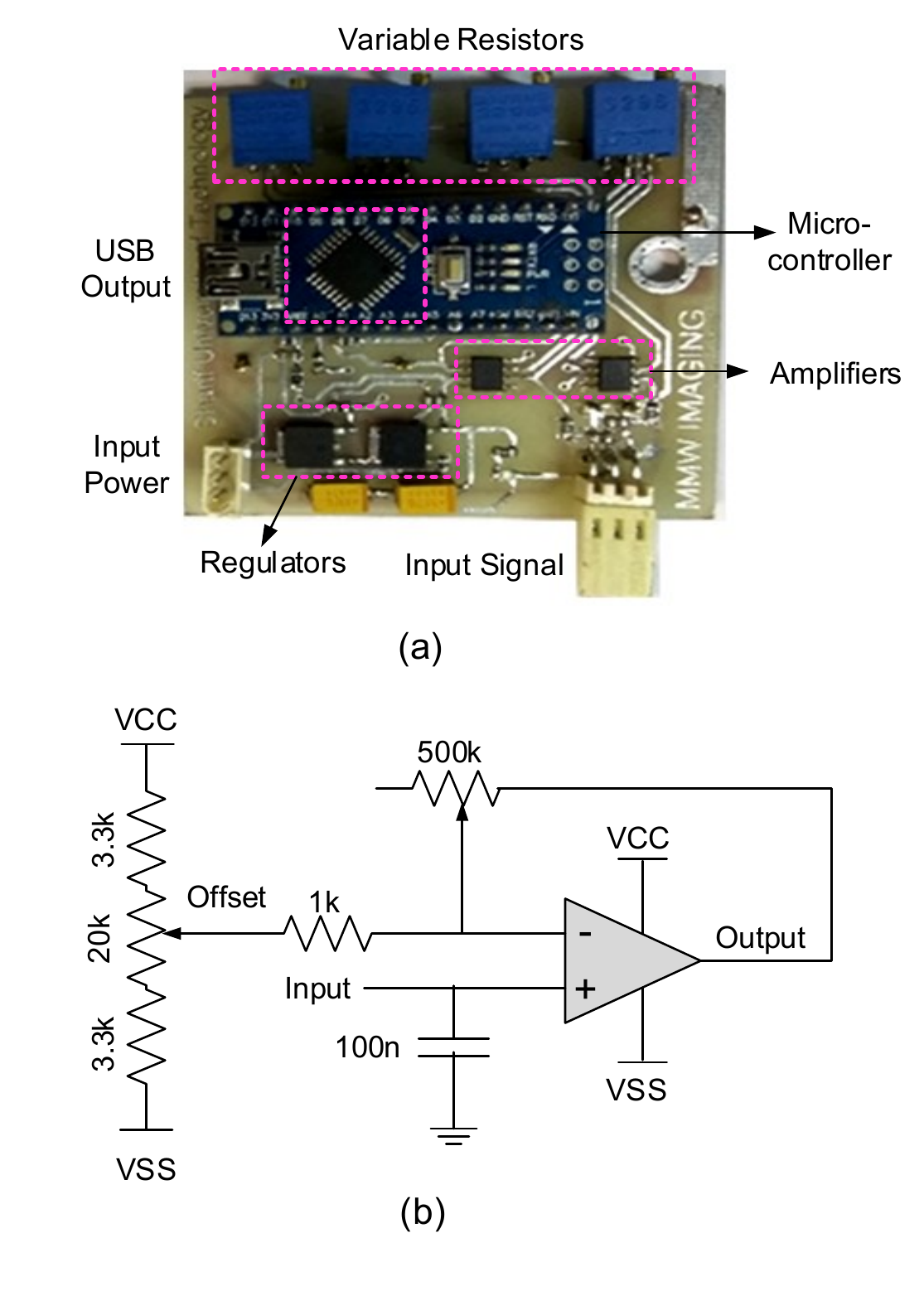}\\
 \caption{Photograph of the data acquisition system. (a) the controller consists of Arduino-nano, LM224D amplifiers, and an offset-removal circuit. (b) Schematic of the off-set removal circuit of our designed controller}\label{controller}
  \end{center}
\end{figure}

To analyze the information received by the antennas, we designed a controller shown in Fig.~\ref{controller}(a). The controller includes an amplifier, an offset-removal circuit and an analog to digital converter (ADC). Fig.~\ref{controller}(b) depicts the schematic of the offset-removal circuit. As observed, it includes a non-inverting operational amplifier (LM224D) and a potentiometer (left panel), used to maximize the amplification. The dynamic range of the output is amplified using an operational amplifier, as long as it is not saturated. The values of the circuit elements are shown in Fig.~\ref{controller}(b). For realizing the ADC, we have used an ATmega328 micro-controller in Arduino-nano. The digital signal is transferred to a computer using USB connectors. The corresponding digital signal is used to reconstruct the image using the GSAFT algorithm described in section II. 
\section{Experimental Results}
\begin{figure}
  \begin{center}
  \includegraphics[width=3.5in]{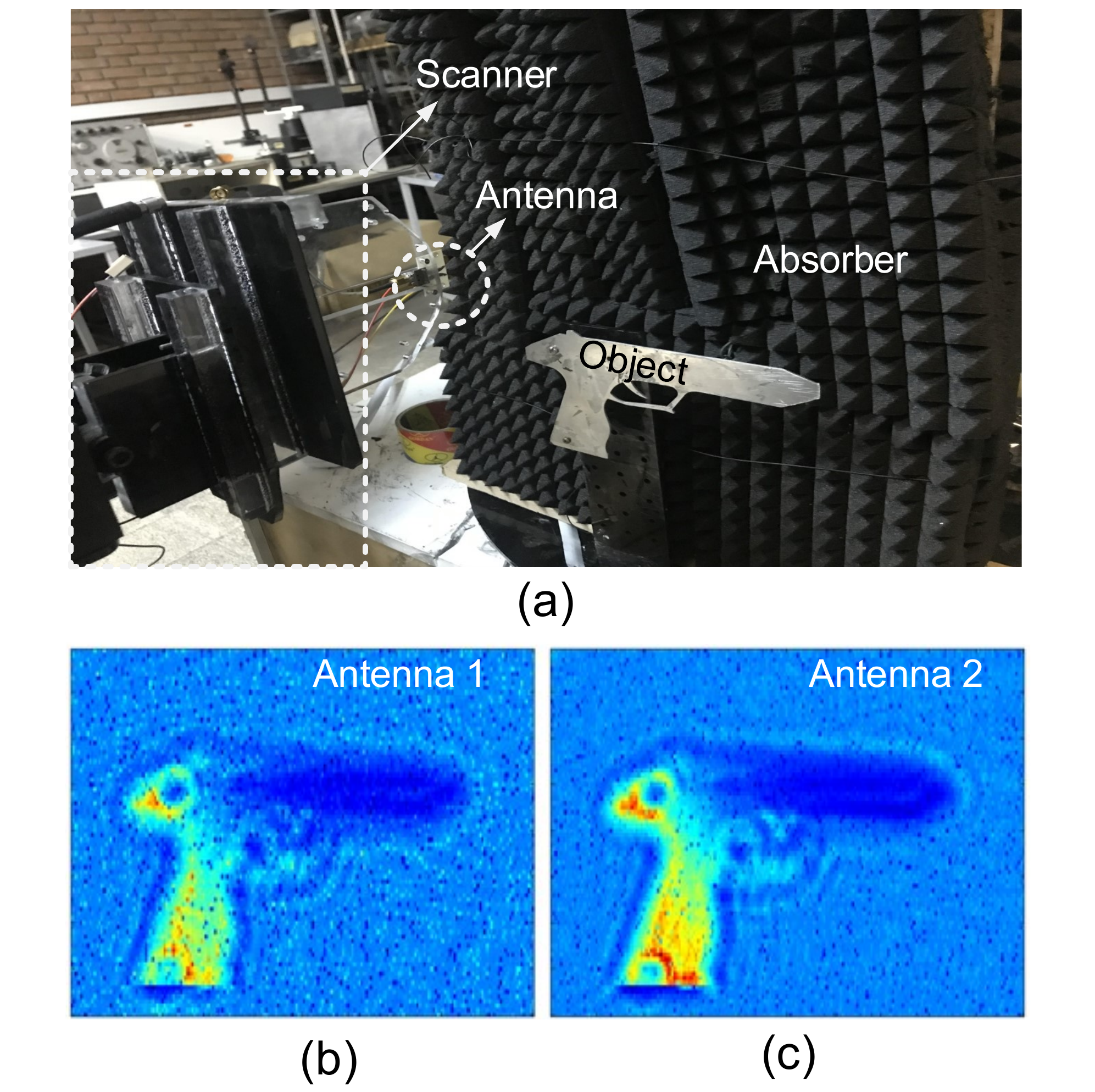}\\
 \caption{Experimental measurements. (a) Photograph of the experimental setup used to perform millimeter wave imaging. (b), (c) Corresponding images of the gun, based on the signal received at the antenna 1 and 2.}\label{result1}
  \end{center}
\end{figure}

To verify the performance of the proposed imaging system, we used a metallic gun model  as the target object.  Fig.~\ref{result1}(a) shows the experimental setup, where the antenna element is raster-scanned over a rectangular plane (note that the center of this rectangular plane is matched with that of the object. The distance between the center of antenna and that of the object is 39\,mm). For mechanical scanning, the step distance along horizontal and vertical directions are 2.5\,mm and 0.5\,mm, respectively. The imaging is performed at a single frequency ($f=26\,GHz$).  Fig.~\ref{result1}(b) and (c) show the corresponding holograms based on the output voltages of the upper and lower diodes, respectively. As it is observed, in both cases, the image of the object is properly reconstructed.\\
\begin{figure}
\centering
\includegraphics[width=3.5in]{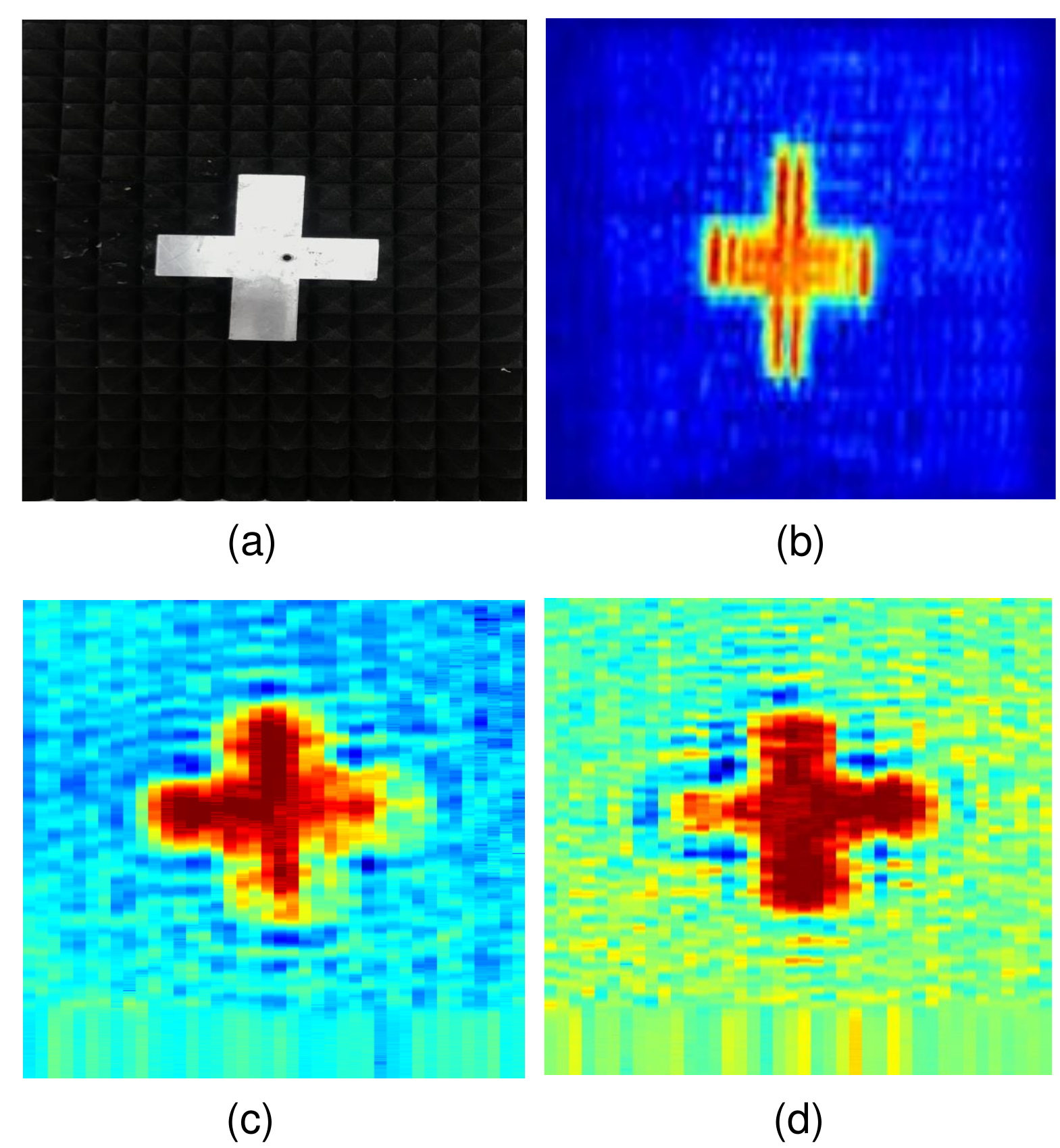}
\caption{Millimeter wave imaging of a metallic cross-shaped object. (a) Photograph of a metallic cross. (b) Reconstructed image of the object.  (c), (d) Corresponding hologram associated with the antenna 1 at the frequency of 26.1\,GHz and 28.1\,GHz, respectively.}
\label{result2}
\end{figure}
In order to perform wideband imaging, we fed the antenna element by a wideband millimeter-wave source in the frequency range of 25.1-30.8\,GHz (the frequency step is 0.5\,GHz). We then performed imaging of a cross-shaped object, whose photograph is shown in Fig.~\ref{result2}(a). Fig.~\ref{result2}(b) shows the corresponding reconstructed image. As observed, the wideband imaging system allows us to reconstruct the sharp edges of the image more precisely. Furthermore, wideband imaging enables focusing on depth locations. This feature is clear from the result of Fig.~\ref{result2}(b), where the reflected field has almost the same amplitude inside the cross area. Insets of Fig.~\ref{result2}(c) ,(d) represent the corresponding hologram associated with the first antenna at two different frequencies, namely 26.1\,GHz (panel c) and 28.1\,GHz (panel d).
\section{Conclusion}
We designed and experimentally demonstrated a wideband millimeter wave imaging system based on SAR technique. By performing measurements, we showed that the proposed imaging system was capable of reconstructing three-dimensional images of arbitrary shaped objects. Compared to traditional microwave imaging systems in which the transmitter and receiver antennas are often printed on two different boards, both transmitter and receiver antennas of our proposed system were printed on the same board. This provides us with a unique opportunity to tackle the problem of phase referencing for different frequencies. The proposed imaging system has potential applications in security scanning, non-destructive sensing, spectroscopy, and medical screening. By substituting the single element antenna of our proposed structure with a linear array, one can benefit from the advantages of electrical scanning.
\section*{Reference}
\bibliographystyle{IEEEtran}
\bibliography{IEEEabrv,Bibliography}
\end{document}